\documentclass{iopart}

\usepackage{iopams}
\usepackage{graphicx}
\usepackage{amssymb}
\usepackage{color}
\usepackage{subfig}
\usepackage{float}
\usepackage{url}

 \expandafter\let\csname equation*\endcsname\relax 
 \expandafter\let\csname endequation*\endcsname\relax 
\usepackage{amsmath}

\begin{document}

\title[Dual-Path Methods for Propagating Quantum Microwaves]{Dual-Path Methods for Propagating Quantum Microwaves}

\author{R Di Candia$^{1,*}$, E P Menzel$^{2,3,*}$, L Zhong$^{2,3}$, F Deppe$^{2,3}$, A Marx$^{2}$, R Gross$^{2,3}$, and E Solano$^{1,4}$ }

\address{$^1$Department of Physical Chemistry, University of the Basque Country UPV/EHU, Apartado 644, E-48080 Bilbao, Spain\\
$^2$Walther-Mei\ss ner-Institut, Bayerische Akademie der Wissenschaften, D-85748 Garching, Germany\\
$^3$Physik-Department, Technische Universit\"at M\"unchen, D-85748 Garching, Germany\\
$^4$IKERBASQUE, Basque Foundation for Science, Alameda Urquijo 36, 48011 Bilbao, Spain \\
\vspace{0.5cm}
$^*$These authors contributed equally to this work.}

\vspace{0.5cm}

\eads{\mailto{rob.dicandia@gmail.com}}

\begin{abstract}
We study quantum state tomography, entanglement detection, and channel noise reconstruction of propagating quantum microwaves via dual-path methods. The presented schemes make use of the following key elements: propagation channels, beam splitters, linear amplifiers, and field quadrature detectors. Remarkably, our methods are tolerant to the ubiquitous noise added to the signals by phase-insensitive microwave amplifiers. Furthermore, we analyze our techniques with numerical examples and experimental data, and compare them with the scheme developed in Eichler $et$ $al$ (2011 $Phys.$ $Rev.$ $Lett.$ {\bf 106} 220503; 2011 $Phys.$ $Rev.$ $Lett.$ {\bf 107} 113601), based on a single path. Our methods provide key toolbox components that may pave the way towards quantum microwave teleportation and communication protocols.
\end{abstract}

\maketitle

\section{Introduction}

In circuit quantum electrodynamics (cQED)~\cite{Blais1, Wallraff1}, a superconducting qubit is coupled to the quantized modes of the electromagnetic field in a superconducting microwave resonator.  In the last few years, we have witnessed tremendous progress in this field, and cQED has developed into a promising platform for quantum computing or quantum communication. Since the typical operating frequency of the system ranges between one and ten gigahertz, and since the output signal of the resonator can propagate along transmission lines, it is of key importance to study how to measure propagating quantum microwave signals and how to use them to perform quantum information processing (QIP) protocols~\cite{Braunstein1}. Two basic problems have to be tackled in order to reach this goal: the measurement of the quantum state with its partial or full reconstruction, known as {\it quantum tomography}, and the detection and quantification of entanglement.

Quantum state tomography is an important tool in quantum information. It consists in the reconstruction of a quantum state by means of measurements. The importance of quantum tomography in QIP lies in its power to benchmark protocols, e.g., it allows to check whether a quantum system evolves as expected. For this reason, great efforts have been taken to develop quantum state tomography techniques and to create suitable measurement devices in various systems. The latter include, e.g., trapped ions~\cite{Wineland1}, optical signals~\cite{Schiller1}, cavity QED~\cite{Solano1, Haroche1} and cQED~\cite{Martinis1}. The resulting progress has also improved the abilities to detect and quantify the fundamental resource of most QIP protocols, i.e., {\it entanglement}. In other words, a direct detection method for entanglement is essential if we want to check the feasibility of such a protocol.

In this article, we describe a quantum state tomography method and a quantum entanglement detection scheme for propagating microwave fields. Despite some theoretical~\cite{Romero1,Peropadre1} and experimental efforts~\cite{Chen1}, no efficient photodetectors are available in the microwave regime yet. Thus, the tomography schemes developed in quantum optics cannot be readily applied to the microwave domain. More specifically, quantum microwaves must be amplified because of their comparably low energy. In practice, this amplification process adds a significant amount of noise~\cite{Menzel1}. The first experimental result of quantum state tomography in the microwave regime is reported by Menzel {\it et al.}~\cite{Menzel1}. It is based on a technique called the {\it dual-path method} (DPM), and was inspired by earlier consideration of Mariantoni {\it et al.}~\cite{Mariantoni1}. Even though the experiment in~\cite{Menzel1} was carried out for coherent states, it involved only few microwave photons and showed that the DPM could be extended to nonclassical states. Such a nonclassical state was finally reconstructed in a later work by Menzel {\it et al.}~\cite{Menzel2}. The first aim of the present paper is to develop the full quantum formalism for the DPM, giving explicit formulas for the quantum state reconstruction. We use only linear devices such as linear detectors, linear amplifiers, and weak thermal or vacuum ancilla states. Furthermore, we avoid standard quantum homodyne detection, which makes use of photodetectors to measure a quadrature observable. Instead, we divide the signal via a \,50:50 beam splitter using the vacuum on the ancillary input port, amplify the output signals by phase-insensitive microwave amplifiers and, finally, measure the field quadratures in each channel.
By calculating suitable auto- and cross- correlations of these noisy signals, we find that it is possible to reconstruct the moments of the input signal and those of the amplifier noise with the knowledge of only the first two moments of the ancilla. The obtained moments can be used to reconstruct the Wigner function of the input signal~\cite{Buzek1}, or to study the behaviour of superconducting devices, e.g., Josephson parametric amplifiers (JPA)~\cite{Yamamoto1, Castellano1, Zhong1}. An alternative approach to tomography of propagating microwaves is homodyne detection by pre-amplifying a single quadrature via a JPA~\cite{Leonhardt1, Mallet1}. In fact the JPA allows, in principle, to measure a single quadrature without adding noise. Since this method is conceptually very different from our moment-based approach, we choose to compare our method to the {\it single-path}, or {\it reference state}, method  (SPM) implemented in~\cite{Eichler1, Eichler2}, which reconstruct the moments of the input signal by using only a single path with a previous tomography of the amplifier noise. Finally, we demonstrate that the dual-path experimental setup can be used to directly detect entanglement between the two output signals of the beam splitter by means of moment based entanglement witnesses (or criteria)~\cite{Vogel1, Miranowicz1, Miranowicz2}. We outline this aspect and describe a scheme to detect these correlations experimentally~\cite{Menzel2}. In summary, we analyze, from a theoretical perspective, the rich physics of quantum microwave signals incident at a beam splitter whose outputs are connected to noisy amplification chains. While most of our results are quite general, we put particular emphasis on the continuous-variable scenario.

The paper is organized in the following way. In section 2, we first present the fundamentals of the dual-path scheme. Then, in section 3, we describe the SPM and the DPM, and compare their performance with respect to the tomography of a few-photon state. In section 4, we discuss the entanglement detection scheme using the dual-path setup. Finally, in section 5, we show experimental examples for quantum tomography of a squeezed state and entanglement detection of a two-mode squeezed state.

\section{Dual-path fundamentals}

In this section, we present the basic elements that we use in the dual-path scheme, providing their input-ouput relations.

\subsection{Beam splitter}

In the quantum regime, a beam splitter is a linear device that superposes two input signals (see Fig.~\ref{elements}(a)). Here, we consider only $50$:$50$ beam splitters with input-output relations 
\begin{align}\label{beamsplitter}
\left(
\begin{matrix}
\hat a_1 \\
\hat a_2
\end{matrix}
\right)=
\frac{1}{\sqrt{2}}
\left(
\begin{matrix}
1 & 1 \\
-1 & 1
\end{matrix}
\right)
\left(
\begin{matrix}
\hat a \\
\hat v
\end{matrix}
\right) ,
\end{align}
where $\hat a$, $\hat v$ are the input signal operators and $\hat a_1$, $\hat a_2$ are the output signal operators, obeying the bosonic commutation relations, e.g.,  $[\hat a,\hat a^\dag]=1$. In the case of propagating signals in the frequency range of $1$-$10$ GHz, hybrid ring structures~\cite{Mariantoni2,Hoffmann1} have already been used as beam splitters in several experiments.

\subsection{Phase-insensitive amplifier}

In our scheme, we use devices that amplify all quadratures in the same way, the so-called phase-insensitive amplifiers (represented in Fig.~\ref{elements}(b)). The input-output relations for these devices are~\cite{Caves1}
\begin{equation}\label{caves}
\hat a'=\sqrt{g}\,\hat a+\sqrt{g-1}\,\hat h_{\text{amp}}^\dag,
\end{equation}
where $\hat h_{\text{amp}}$ is a bosonic operator needed in order to make $\hat a'$ to fulfill the bosonic commutation relations and $g>1$ is the power gain. An example for devices working in our frequency range and providing sufficient gain are high-electron-mobility transistor (HEMT) amplifiers. In these devices, the amount of noise added depends on the design, material system and operation temperature, but $\hat h_{\text{amp}}$ is typically associated with a state with $10$--$20$ photons.

\subsection{IQ-mixer}

In quantum optics, a quadrature measurement can be implemented via a homodyne technique using two photodetectors. This method consists in superimposing the signal with a strong coherent state on a $50$:$50$ beamsplitter, measuring the number of photons at the outputs and calculating their difference. 
\begin{figure}[t]
\centering
\includegraphics[width=\textwidth]{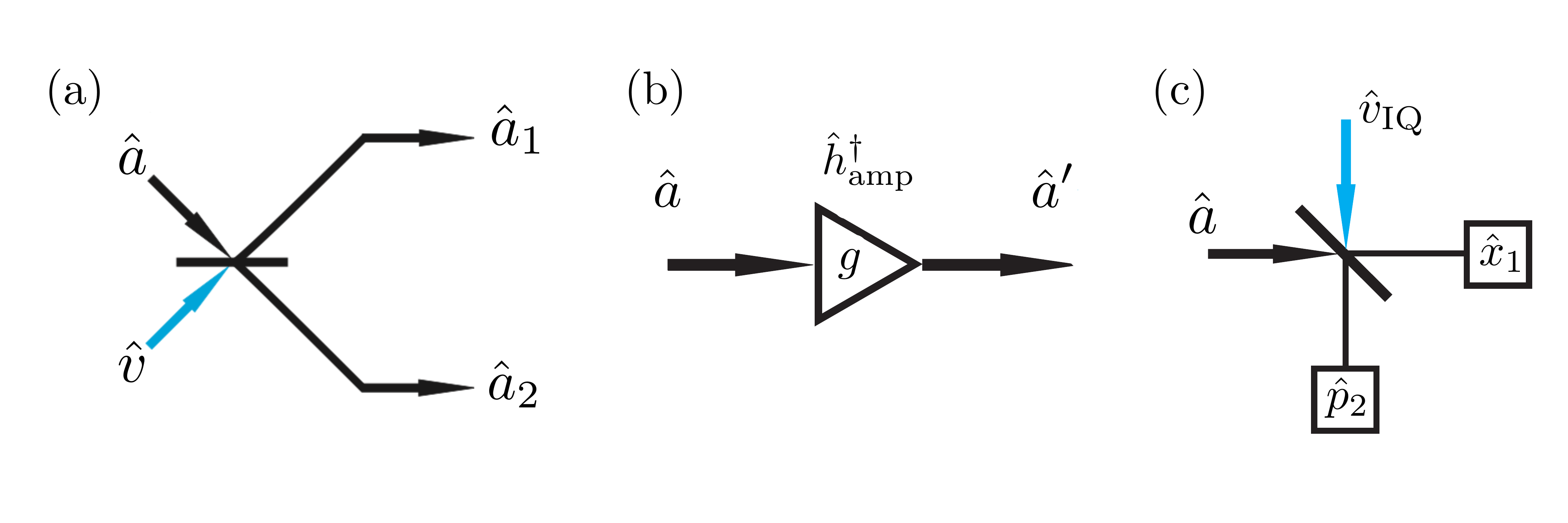}
\caption{(a) Beam splitter scheme. (b) Phase-insensitive linear amplifier scheme. (c) Beam splitter model for the IQ mixer.}\label{elements}
\end{figure}
It involves photon-number measurements, which can be implemented in quantum optics using photodetectors, and it allows for single shot quadrature measurements. In the microwave regime, efficient, low-noise photodetectors are not readily available. However, if the signal is strong enough, we can measure its quadratures using an in-phase-quadrature (IQ) mixer~\cite{Collins1}. The IQ mixer allows us to measure simultaneously both quadratures at the cost of adding noise. It can be modelled using a beam splitter and an ancilla $\hat v_{\text{IQ}}$ in a thermal state~\cite{daSilva1}, e.g., the vacuum (see Fig.~\ref{elements}(c)). From the input-output relations~\eqref{beamsplitter}, we see that the quadrature operators $\hat X_1\equiv\sqrt{2}\hat x_{1}$ and $\hat P_2\equiv\sqrt{2}\hat p_{2}$ (where $1$ and $2$ label the two outputs of the beam splitter) have the same first moment of $\hat x_a$ and $\hat p_a$ respectively, but different variances:
\begin{align}
\langle\hat X_1\rangle&=\langle\hat x_a\rangle,\quad \quad\Delta \hat X_1^2=\Delta \hat x_{a}^2+\Delta \hat x_{v_{\text{IQ}}}^2\\
\langle\hat P_2\rangle&=\langle\hat p_a\rangle,\quad\quad \Delta \hat P_2^2=\Delta \hat p_{a}^2+\Delta \hat p_{v_{\text{IQ}}}^2.
\end{align}
Here, $\Delta \hat A^2\equiv \langle\hat A^2\rangle-\langle\hat A\rangle^2$ indicates the variance of the observable $\hat A$, and we have defined the quadratures as $\hat x\equiv(\hat a+\hat a^\dag)/\sqrt{2}$ and $\hat p\equiv-i(\hat a-\hat a^\dag)/\sqrt{2}$.

\section{Quantum tomography of the input state}

In this section, we describe the SPM~\cite{Eichler1, Eichler2}, and the quantum DPM~\cite{Menzel1, Menzel2}. We test the performance of the two techniques by numerical simulations applying a weak squeezed state. 

\subsection{Single-path (reference-state) method revisited}\label{referencestate}

\begin{figure}[t]
\centering

\includegraphics[width=0.85\textwidth]{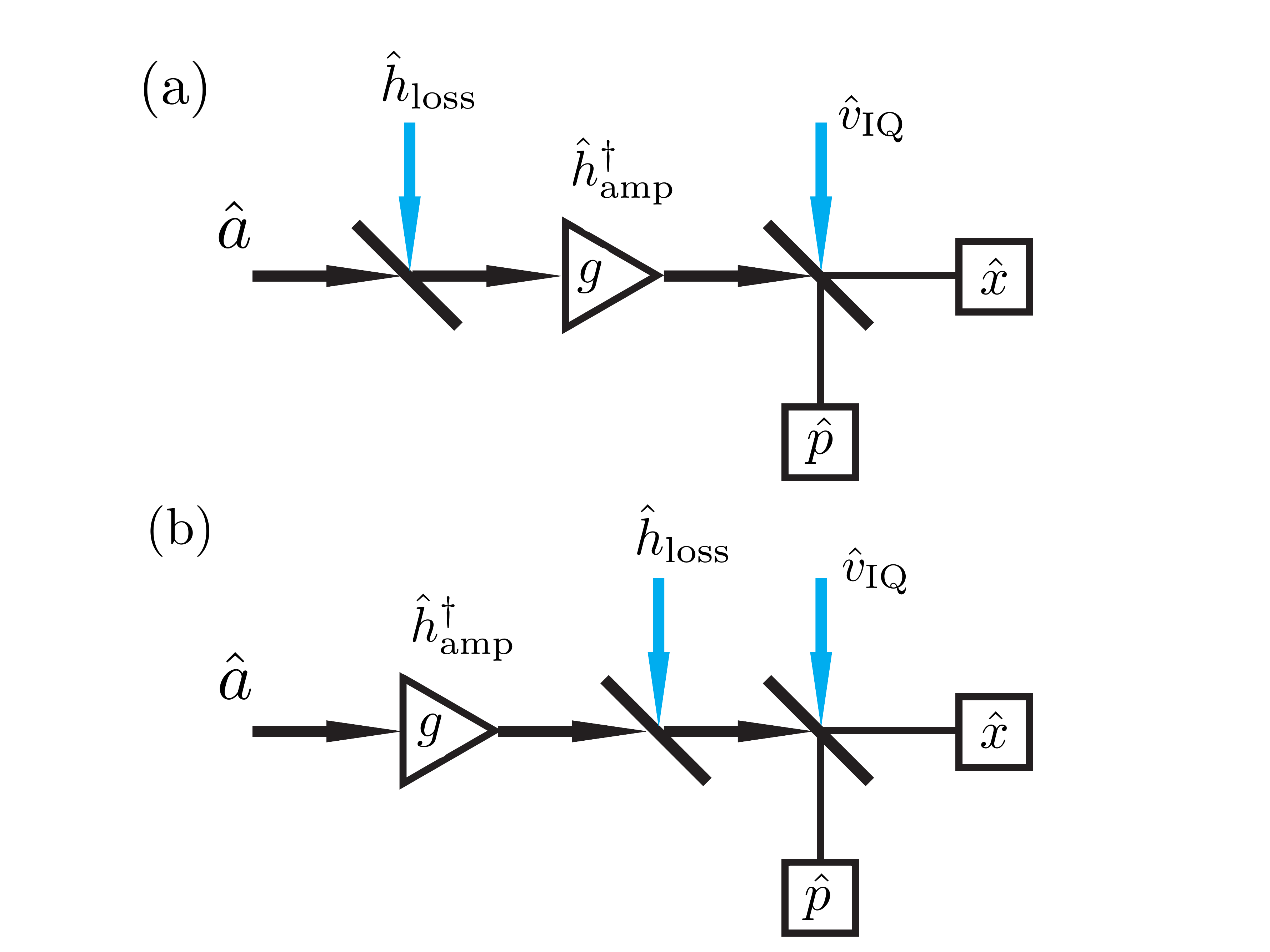}
\caption{Single-path scheme with losses and amplifier in series. (a) Loss before amplification. Here, $\hat h_{\text{loss}}$  is the noise added due to the losses, $\hat h_{\text{amp}}$ is the noise added by the amplifier, $g$ is the power gain and $\hat v_{\text{IQ}}$ is the ancilla needed to measure $\hat x$ and $\hat p$ (see IQ mixer model). (b) Loss after amplification.} \label{one-path}
\end{figure}

Suppose we have a microwave signal $\hat a$ and we want to know its state. As the interesting signals are too weak to be detected, we need to amplify before performing a measurement, as in Eq.~\eqref{caves} (see Fig.~\ref{one-path}).
In the SPM, we first send a known state to characterize $\hat h_{\text{amp}}$, and then we measure with $\hat a$ as input. If we send, for instance, a known coherent state $|\alpha\rangle$ as input, we can retrieve the moments of $\hat h_{\text{amp}}$ from the moments of the outcoming signals in a recursive way
\begin{align}\label{ampmoments}
\langle\hat h_{\text{amp}}^{ l}\hat h_{\text{amp}}^{\dag m}\rangle&=\frac{\langle\hat a'^{\dag l}\hat a'^{m}\rangle}{(g-1)^{\frac{l+m}{2}}}-\sum_{i_1=0}^l\sum_{i_2=0}^{m-1}\binom{l}{i_1}\binom{m}{i_2}\left(\frac{g}{g-1}\right)^{\frac{l+m-(i_1+i_2)}{2}}(\alpha^*)^{l-i_1}\alpha^{m-i_2}\nonumber \\
\quad& \times\langle\hat h_{\text{amp}}^{ i_1}\hat h_{\text{amp}}^{\dag i_2}\rangle-\sum_{i_1=0}^{l-1}\binom{l}{i_1}\left(\frac{g}{g-1}\right)^{\frac{l-i_1}{2}}(\alpha^*)^{l-i_1}\langle\hat h_{\text{amp}}^{ i_1}\hat h_{\text{amp}}^{\dag m}\rangle.
\end{align}
The gain $g$ can, in principle, be found by comparing the first moments ($g=|\langle\hat a'\rangle/\alpha|^2$, assuming $\alpha\not=0$). However, in reality this is difficult since in a typical experiment situation $\alpha$ is not well known. Hence, $g$ is determined in a calibration experiment with thermal states~\cite{Menzel2, Mariantoni2}. Once the calibration is done, usually the vacuum ($\alpha=0$) is chosen as reference state. This leads to the formula
\begin{equation}\label{ampmoments1}
\langle\hat h_{\text{amp}}^{ l}\hat h_{\text{amp}}^{\dag m}\rangle=\frac{\langle\hat a'^{\dag l}\hat a'^{m}\rangle}{(g-1)^{\frac{l+m}{2}}}.
\end{equation}

Once we have characterized the ampflier noise, we can perform tomography of a general quantum state $\hat a$. In fact, by inverting Eq.~\eqref{caves}, we obtain
\begin{align}\label{inv}
\langle\hat a^{ \dag l}\hat a^{ m}\rangle&=\frac{\langle\hat a'^{\dag l}\hat a'^{m}\rangle}{g^{\frac{l+m}{2}}}-\sum_{i_1=0}^l\sum_{i_2=0}^{m-1}\binom{l}{i_1}\binom{m}{i_2}\left(\frac{g-1}{g}\right)^{\frac{l+m-(i_1+i_2)}{2}}\langle\hat a^{\dag i_1}\hat a^{i_2}\rangle\langle\hat h_{\text{amp}}^{ l-i_1}\hat h_{\text{amp}}^{\dag m-i_2}\rangle\nonumber \\
\quad&-\sum_{i_1=0}^{l-1}\binom{l}{i_1}\left(\frac{g-1}{g}\right)^{\frac{l-i_1}{2}}\langle\hat a^{\dag i_1}\hat a^m\rangle\langle\hat h_{\text{amp}}^{ l-i_1}\rangle,
\end{align}
where $\langle\hat h_{\text{ampl}}^{ l}\hat  h_{\text{ampl}}^{\dag m}\rangle$ is derived in Eq.~\eqref{ampmoments}.

In an experiment, the channel is lossy. We can model the losses  with a beam splitter:
\begin{equation}\label{losses}
\hat b_{\text{out}}=\sqrt{\eta} \,\hat b_{\text{in}} +\sqrt{1-\eta}\,\hat h^{}_{\text{loss}},
\end{equation}
where $0<\eta<1$ is the power loss, and $\hat h^{}_{\text{loss}}$ is the noise added by the environment, strictly related to its temperature.
If we have losses and amplification in series as in Fig.~\ref{one-path}, we can consider the channel as amplifying or lossy, depending on the value of $g\eta$. For the detection of quantum microwaves, a large value of the effective gain $g\eta$ on the order of $10^4$ is required, so that the amplified signal amplitudes are well above the noise added by the next components. First, we consider the case that the losses occur before the signal is amplified (see Fig.~\ref{one-path}(a)) and rewrite Eq.~\eqref{caves} as
\begin{equation}\label{noiseampl}
\hat a'=\sqrt{g\eta}\hat a+\sqrt{g\eta-1}\hat V^\dag,
\end{equation}
with $g\eta>1$ and $\hat V=\sqrt{\frac{1-\eta}{\eta-\frac{1}{g}}}\hat h_{\text{loss}}^\dag+\sqrt{\frac{1-\frac{1}{g}}{\eta-\frac{1}{g}}}\hat h^{}_{\text{amp}}\simeq \sqrt{\frac{1}{\eta}-1}\hat h_{\text{loss}}^\dag+\sqrt{\frac{1}{\eta}}\hat h_{\text{amp}}\equiv \hat V^{}_{\text a}$ in the limit $g\gg1$, obtaining an equivalent model, but with an effective gain and noise mode $\hat V$. Therefore, we can still use Eq.~\eqref{ampmoments} to reconstruct the moments of $\hat V$ and Eq.~\eqref{inv} to retrieve the moments of $\hat a$.
Second, if the losses occur after the amplification (see Fig.~\ref{one-path}(b)), we get the same effective gain $g\eta$, but the noise mode $\hat V=\sqrt{\frac{1-\frac{1}{g}}{1-\frac{1}{g\eta}}}\hat h^{}_{\text{amp}}+\sqrt{\frac{1-\eta}{g\eta-1}}\hat h_{\text{loss}}^\dag\simeq \hat h^{}_{\text{amp}}\equiv \hat V^{}_{\text b}$ in the limit $g\eta\gg1$.
We conclude that the lossy models are equivalent to the ideal one, but with different amplifier gains. Furthermore, we note that if the gain is large enough, the effect of the losses occurring after the amplification are negligible. In classical network theory, this is well-known~\cite{Friis}. In fact, assuming $\hat h^{}_{\text{amp}}$ and $\hat h^{}_{\text{loss}}$ in thermal states and $g\gg1$, we have that $\langle\hat V_{\text b}^\dag\hat V_{\text b}\rangle=\langle\hat h_{\text{amp}}^{\dag}\hat h_{\text{amp}}^{}\rangle<(\frac{1}{\eta}-1)+(\frac{1}{\eta}-1)\langle\hat h_{\text{loss}}^\dag\hat h^{}_{\text{loss}}\rangle+\frac{1}{\eta}\langle\hat h_{\text{amp}}^\dag\hat h^{}_{\text{amp}}\rangle=\langle\hat V_{\text a}^\dag \hat V^{}_{\text a}\rangle$, $\forall \eta \in (0,1)$, so the model in Fig.~\ref{one-path}(b) contains less noise than the one in Fig.~\ref{one-path}(a). Regarding the optimization of experimental setups, minimizing the losses occurring before amplification is crucial under the assumption that $g$ is large enough, which usually holds. In the remainder of this work, we use the ideal amplifier model of~\eqref{caves}, treating $g$ and $\hat  h_{\text{amp}}$ as effective parameters.

\subsection{Dual-path method}

\begin{figure}[t]
\centering
\includegraphics[width=9 cm]{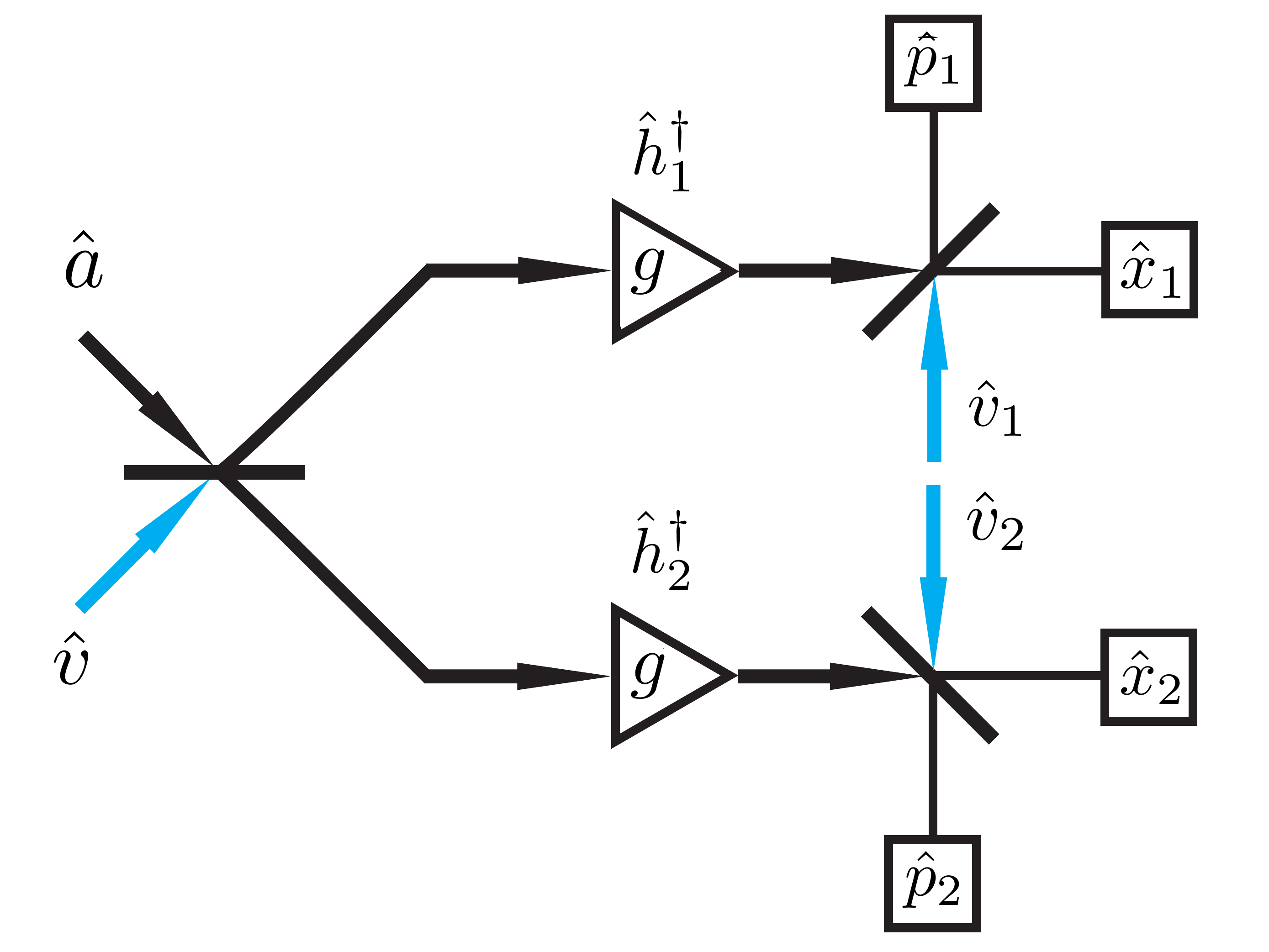}
\caption{Dual-path scheme for the quantum state tomography of a microwave signal. The signal is divided into two parts via a beam splitter. The two outputs are then amplified and detected. The auto- and cross-correlations of the two measured quadratures allow for the retrieval of both the moments of the input signal and those of the amplifier noise fields.} \label{dualpath}
\end{figure}

We now describe the quantum formalism of the DPM, including explicit formulas for the reconstruction of the moments. Our method assumes minimum information on an ancilla, i.e., the knowledge of its first two moments, and it allows to reconstruct the moments of the input signal and of the amplifier noise fields at the same time. The scheme is depicted in Fig.~\ref{dualpath}. We send a signal $\hat a$ to a 50:50 beam splitter with an ancilla $\hat v$, and we direct the output fields ($\hat c_1$ and $\hat c_2$) of the beam splitter to two amplifiers, obtaining
\begin{align}
\hat c'_1&=\frac{\sqrt{g}}{\sqrt{2}}(\hat a+\hat v)+\sqrt{g-1}\hat h_{1}^\dag, \label{c1}\\
\hat c'_2&=\frac{\sqrt{g}}{\sqrt{2}}(-\hat a+\hat v)+\sqrt{g-1}\hat h_{2}^\dag, \label{c2}
\end{align}
where we have assumed for simplicity the same gain $g$ for both channels. Lastly, we measure the quadratures $\hat x$ and $\hat p$ of both channels with IQ mixers. It is useful to introduce the complex envelopes~\cite{daSilva1} 
\begin{equation}\label{envelope}
\hat S_{1,2}\equiv \frac{\hat x^{}_{1,2}+i\hat p_{1,2}^{}}{\sqrt{g}}=\frac{\hat c'_{1,2}+\hat v_{1,2}^\dag}{\sqrt{g}},
\end{equation}
where $\hat x_{1,2}^{}$ and $\hat p_{1,2}^{}$ are the quadrature operators for the channels 1 and 2, $\hat v_{1,2}$ are the ancillas in the IQ mixer model, and the last equality can be easily checked from the quadrature definition and the beam splitter relation~\cite{daSilva1}. Finally, we obtain 
\begin{equation}\label{S2}
\hat S_1=\frac{\hat a+\hat v}{\sqrt{2}}+\hat V_1^\dag,\quad\quad \hat S_2=\frac{-\hat a+\hat v}{\sqrt{2}}+\hat V_2^\dag,
\end{equation}
with $\hat V_{1,2}=\sqrt{1-\frac{1}{g}}\hat h_{1,2}+\sqrt{\frac{1}{g}}\hat v_{1,2}\simeq\hat h_{1,2}$ in the limit $g\gg1$. 

We now iteratively find the relation between the moments of the detected signals and the moments of the input signal. For the first moment, we obtain
\begin{equation}
\langle\hat S_1\rangle=\frac{\langle\hat a\rangle+\langle\hat v\rangle}{\sqrt{2}},\quad\langle\hat S_2\rangle=\frac{-\langle\hat a\rangle+\langle\hat v\rangle}{\sqrt{2}}, \label{first2}
\end{equation}
where we have assumed $\langle\hat V_1^\dag\rangle=\langle\hat V_2^\dag\rangle=0$. This is a safe assumption for any kind of noise. From~\eqref{first2} we derive
\begin{align}
&\langle\hat a\rangle=\frac{1}{\sqrt{2}}(\langle\hat S_1\rangle-\langle\hat S_2\rangle),\\
&\langle\hat v\rangle=\frac{1}{\sqrt{2}}(\langle\hat S_1\rangle+\langle\hat S_2\rangle). \label{firstancilla}
\end{align}
We assume the knowledge of the first two moments of the ancilla $\hat v$, so we leave~\eqref{firstancilla} as a consistency check.
From the second moments
\begin{align}
&\langle\hat S_1^2\rangle=\frac{\langle\hat a^2\rangle}{2}+\frac{\langle\hat v^2\rangle}{2}+\langle\hat a\rangle\langle\hat v\rangle+\langle\hat V_1^{\dag 2}\rangle,\\
&\langle\hat S_2^2\rangle=\frac{\langle\hat a^2\rangle}{2}+\frac{\langle\hat v^2\rangle}{2}-\langle\hat a\rangle\langle\hat v\rangle+\langle\hat V_2^{\dag 2}\rangle,\\
&\langle\hat S_1\hat S_2\rangle=-\frac{\langle\hat a^2\rangle}{2}+\frac{\langle\hat v^2\rangle}{2}, \label{cross}
\end{align}
we get by linear inversion
\begin{align}
&\langle\hat a^2\rangle=-2\langle\hat S_1\hat S_2\rangle+\langle\hat v^2\rangle,\\
&\langle\hat V_1^{\dag 2}\rangle=\langle\hat S_1^2\rangle-\frac{\langle\hat a^2\rangle}{2}-\frac{\langle\hat v^2\rangle}{2}-\langle\hat a\rangle\langle\hat v\rangle,\\
&\langle\hat V_2^{\dag 2}\rangle=\langle\hat S_2^2\rangle-\frac{\langle\hat a^2\rangle}{2}-\frac{\langle\hat v^2\rangle}{2}+\langle\hat a\rangle\langle\hat v\rangle.
\end{align}
Similar formulas hold for $\langle\hat a^\dag \hat a\rangle$ and $\langle\hat V_{1,2}\hat V_{1,2}^\dag\rangle$ (see Appendix). In the derivation of the above formulas, the mutual statistical independence of $\hat a$, $\hat v$, $\hat V_1$ and $\hat V_2$ is crucial. This is a reasonable assumption if we consider, for instance, that the two amplifiers are spatially well separated and, hence, have different environments.  

We note that we have derived the moments of the input field via the cross-correlation terms, and, using this result, we have derived the moments of the noise via the auto-correlation terms. This calculation can be iterated over every order, and in this way, we can retrieve the moments of the input state and the noise. We also note that from the third moment on, we do not need further assumptions on the ancilla $\hat v$. For instance,  the third moments of the measurement outcomes are
\begin{align}
&\langle \hat S_1^3\rangle=\frac{\langle\hat a^3\rangle}{2\sqrt{2}}+\frac{\langle\hat v^3\rangle}{2\sqrt{2}}+\langle\hat V_1^{\dag3}\rangle+\langle f_{\hat S_1^3 }(\hat a,\hat v, \hat V_1)\rangle,\\
&\langle\hat S_1^2\hat S_2\rangle=-\frac{\langle\hat a^3\rangle}{2\sqrt{2}}+\frac{\langle\hat v^3\rangle}{2\sqrt{2}}+\langle f_{\hat S_1^2\hat S_2 }(\hat a,\hat v, \hat V_1,\hat V_2)\rangle,\\
&\langle\hat S_1\hat S_2^2\rangle=\frac{\langle\hat a^3\rangle}{2\sqrt{2}}+\frac{\langle\hat v^3\rangle}{2\sqrt{2}}+\langle f_{\hat S_1\hat S_2^2 }(\hat a,\hat v, \hat V_1,\hat V_2)\rangle,\\
&\langle \hat S_2^3\rangle=-\frac{\langle\hat a^3\rangle}{2\sqrt{2}}+\frac{\langle\hat v^3\rangle}{2\sqrt{2}}+\langle\hat V_2^{\dag3}\rangle+\langle f_{\hat S_2^3 }(\hat a,\hat v, \hat V_2)\rangle,
\end{align}
where the $f_{\dots}$ are functions of the operators, and $\langle f_{\dots}\rangle$ contain moments up to the second order. Making the right combinations, one derives
\begin{align}
&\langle\hat a^3\rangle=\sqrt{2}(\langle\hat S_1\hat S_2^2\rangle-\langle\hat S_1^2\hat S_2\rangle+\langle f_{\hat S_1^2\hat S_2}\rangle-\langle f_{\hat S_1\hat S_2^2 }\rangle),\\
&\langle\hat v^3\rangle=\sqrt{2}(\langle\hat S_1\hat S_2^2\rangle+\langle\hat S_1^2\hat S_2\rangle-\langle f_{\hat S_1^2\hat S_2}\rangle-\langle f_{\hat S_1\hat S_2^2 }\rangle),\\
&\langle\hat V_1^{\dag3}\rangle=\langle \hat S_1^3\rangle-\frac{\langle\hat a^3\rangle}{2\sqrt{2}}-\frac{\langle\hat v^3\rangle}{2\sqrt{2}}-\langle f_{\hat S_1^3 }\rangle,\\
&\langle\hat V_2^{\dag3}\rangle=\langle \hat S_2^3\rangle+\frac{\langle\hat a^3\rangle}{2\sqrt{2}}-\frac{\langle\hat v^3\rangle}{2\sqrt{2}}-\langle f_{\hat S_2^3 }\rangle.
\end{align}
This calculation can be generalized to any higher-order moments (see Appendix). Summarizing, we present a method to simultaneously reconstruct the moments of an input state and the moments of the noise added in the channels, assuming a minimum information on an ancilla.

\subsection{Comparison of the SPM and the DPM}\label{comparison}

In this section, we compare the DPM to the SPM with respect to the reconstruction of the quantum state moments of an input signal. We numerically test the methods with a squeezed state with squeezing parameter $\xi=0.5i$~\cite{Barnett} as input state, corresponding to a photon number of $0.27$. We model the amplifier noise fields and the ancilla states with thermal states with photon numbers $n_{\text{amp}}$ and $n_{\text{anc}}$ and restrict our analysis to the first four moments. We have simulated several experiments to study the dependence of the statistical uncertainty of the reconstructed signal moments on the number of measurements $N$ and on the photon number of the amplifier noise $n_{\text{amp}}$. For fixed $n_{\text{amp}}$, we compare the two methods in the following way. For the SPM we perform $N$ measurements allowing the reconstruction of the amplifier noise by using a vacuum state as input. 
Then, we simulate $N$ measurement results for the reconstruction of the squeezed state chosen as input. For the DPM, instead, we simulate $N$ measurements of each channel, that allow us to calculate all the needed auto- and cross-correlations to reconstruct the input state.  To obtain the individual measurements at the outputs of the IQ mixers, we use random numbers obeying a multivariate Gaussian statistics. The latter is determined by a covariance matrix, which is calculated using the input output relations of all components. For a certain $N$, we estimate the average values of the simulated quadrature moments by using the sample mean. For the DPM the sample mean is
\begin{equation}
\langle {x'}_1^{k_1}{p'}_1^{l_1}{x'}_2^{k_2}{p'}_2^{l_2}\rangle_{est}=\frac{1}{N}\sum_{j=1}^N{x'}_{1j}^{k_1}{p'}_{1j}^{l_1}{x'}_{2j}^{k_2}{p'}_{2j}^{l_2},
\end{equation}
where $j$ labels the single outcomes. For the SPM we apply an analogous formula. We use the reconstruction formulas of the SPM and DPM to retrieve the moments of the input states. In order to access the statistical properties of the two methods, the described procedure is repeated $5000$ times (for each parameter combination). We divide the data into $20$ blocks with $250$ samples each. The $250$ samples are used to calculate the standard deviation $\sigma$ of the reconstructed signal moments for the DPM and SPM. Finally, we compare the two methods by calculating the ratio $\sigma_{\text{DP}}/\sigma_{\text{SP}}$. Thereby, a higher standard deviation means worse performance. The $20$ blocks allow us to estimate the uncertainty of the obtained mean value of the ratio $\sigma_{\text{DP}}/\sigma_{\text{SP}}$.  

\begin{figure}[t]
\centering
\includegraphics[width=9 cm]{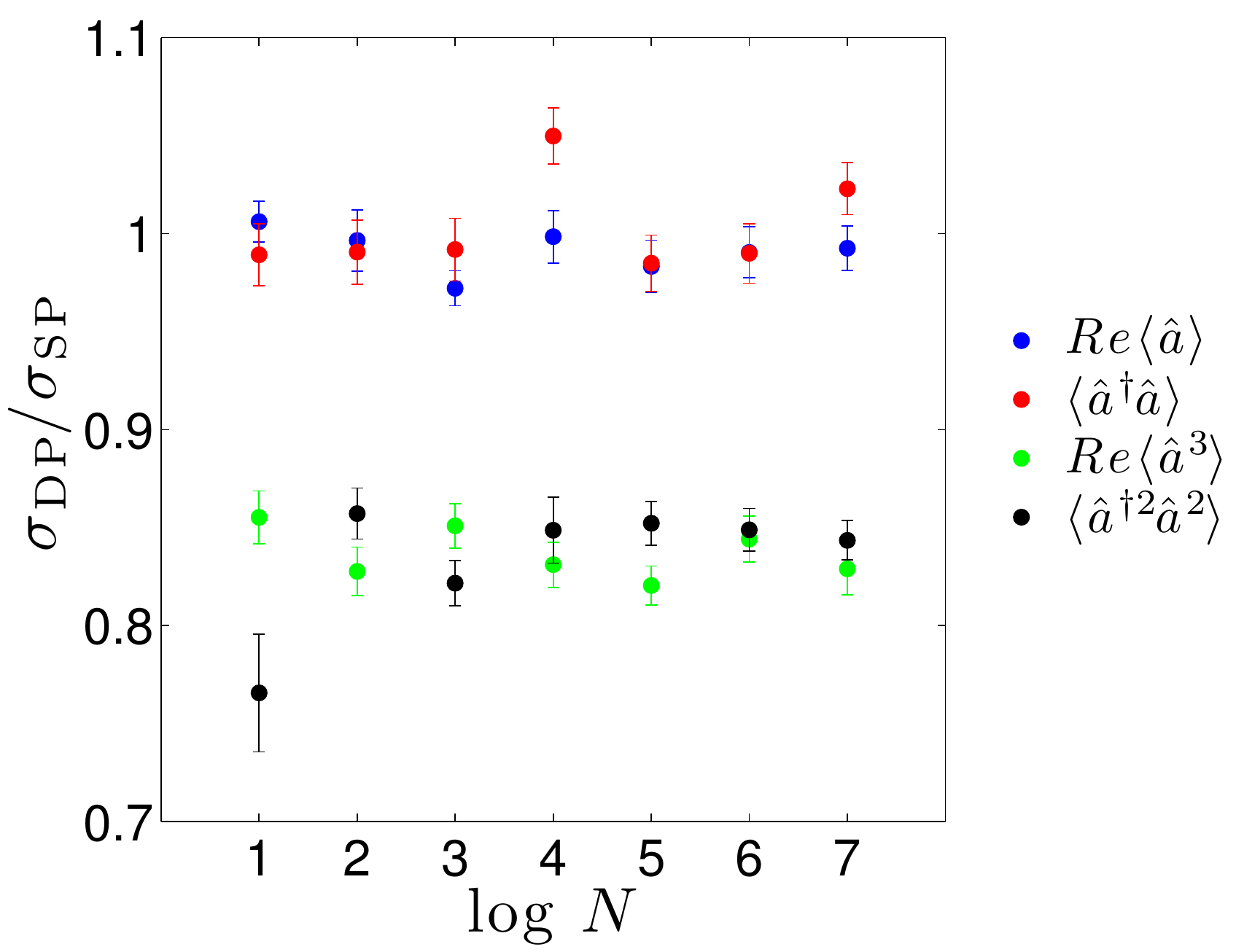}
\caption{Ratio of the standard deviations for (selected) reconstructed signal moments of the DPM ($\sigma_{\text{DP}}$) and the SPM ($\sigma_{\text{SP}}$) in dependence on the number of measurements ($N$).} \label{Figure1}
\end{figure}
\begin{figure}[H]
\centering
\includegraphics[width=\textwidth]{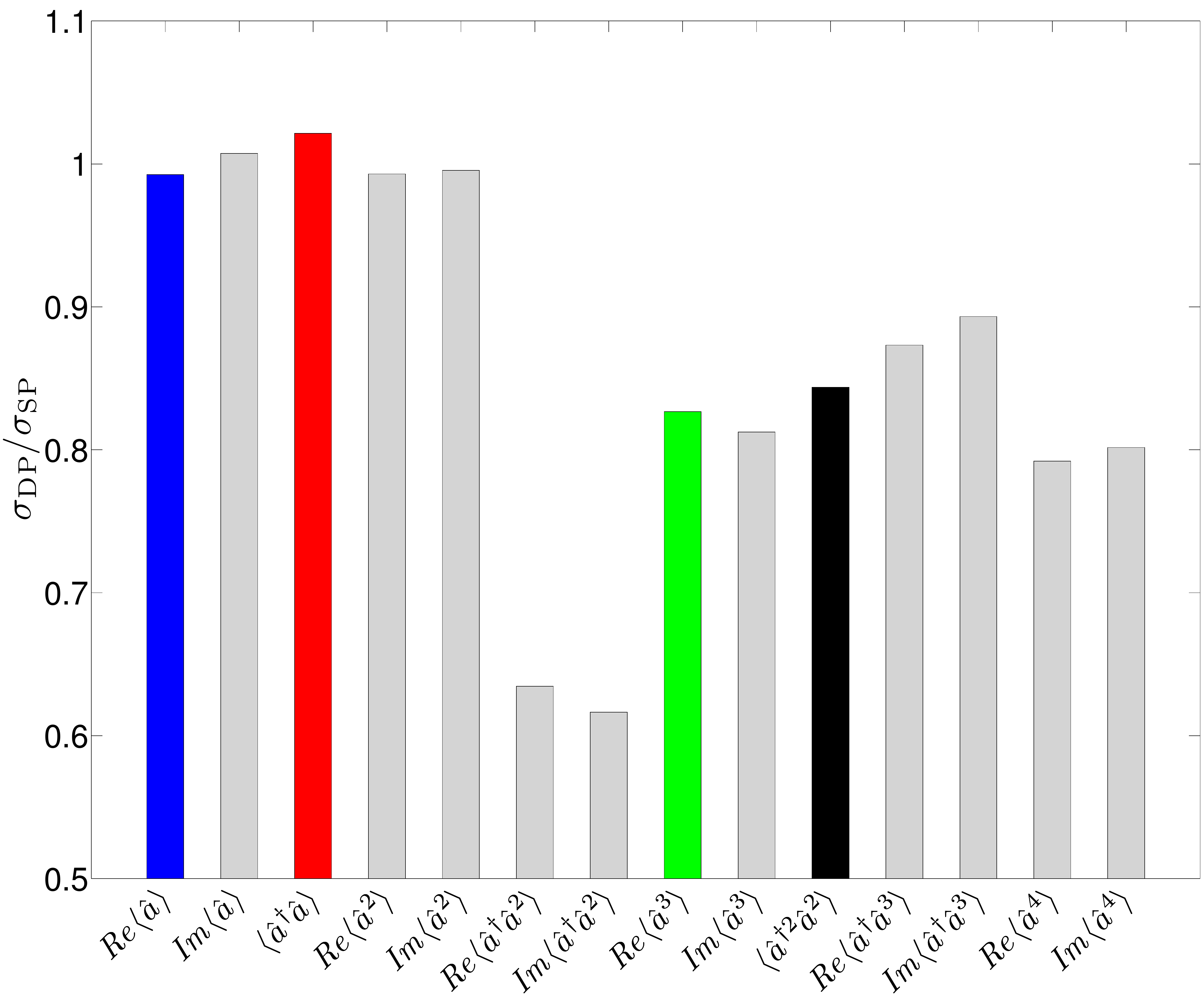}
\caption{Ratio of the standard deviations for the reconstruction of the DPM ($\sigma_{\text{DP}}$) and the SPM ($\sigma_{\text{SP}}$) for all moments until the fourth order.}\label{Figure2}
\end{figure}
\begin{figure}[t]
\centering
\includegraphics[width=8 cm]{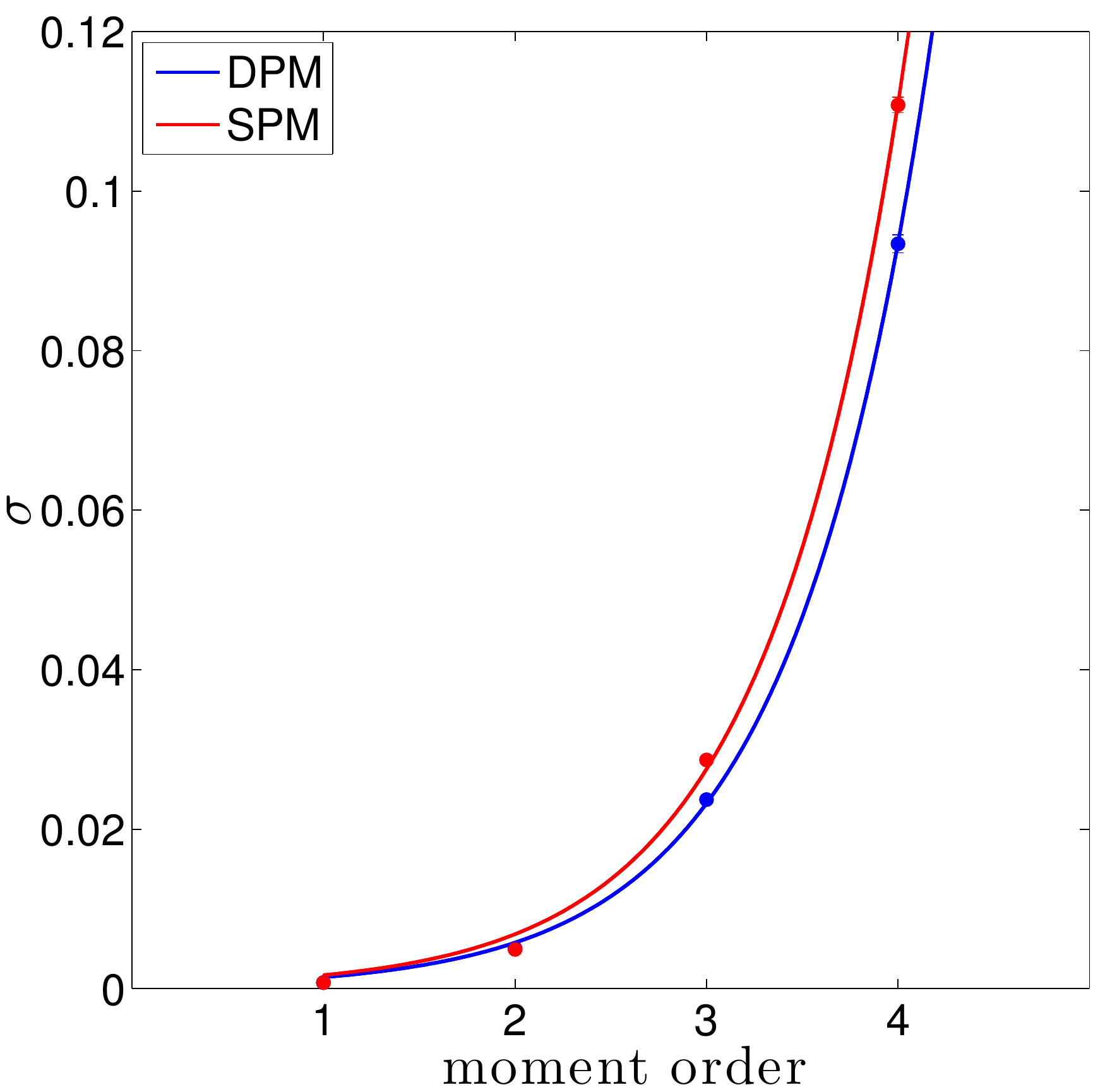}
\caption{Standard deviations ($\sigma$) for the reconstruction using the DPM and the SPM for different moment orders. Symbols: simulation results. Solid lines: exponential fit curves.} \label{Figure3}
\end{figure}
\noindent In experiments at millikelvin temperatures $n_{\text{anc}}$ is negligible small. Therefore, we set $n_{\text{anc}}=0$ in all presented simulations. The results are depicted in Figures~\ref{Figure1},~\ref{Figure2} and~\ref{Figure3}, where we assume $n_{\text{amp}}=10$ for all amplifiers. Additionally, we have assumed that the first moment of the noise is zero in the SPM. This is not needed in the single-path reconstruction, but it is necessary if we want to have a fair comparison to the DPM, as in the latter this assumption is fundamental. In Figures~\ref{Figure1},~\ref{Figure2}, and~\ref{Figure3}, we see that the DPM and the SPM have the same performance regarding the first and second moments, while for the third and the fourth moments the dual-path has an edge. The result is not dependent on the number of measurements (see Fig.~\ref{Figure1}).  For an individual moment, the standard deviation scales as $1/\sqrt{N}$ (data not shown).
In Figure~\ref{Figure2} we note that the moment $\langle\hat a^\dag_{}a^2_{}\rangle$ performs better than $\langle\hat a^{3}_{}\rangle$ even if they are of the same order. This is because $\langle\hat a^\dag_{}a^2_{}\rangle$ can be estimated in more ways than $\langle\hat a^{3}_{}\rangle$, and we are considering the average over all the possible combinations (see Appendix). Figure~\ref{Figure3} shows that for both methods the standard deviation scales exponentially with the moment order, but the rate of the DPM is lower.

\begin{figure}[H]
\centering
\includegraphics[width=1.05\textwidth]{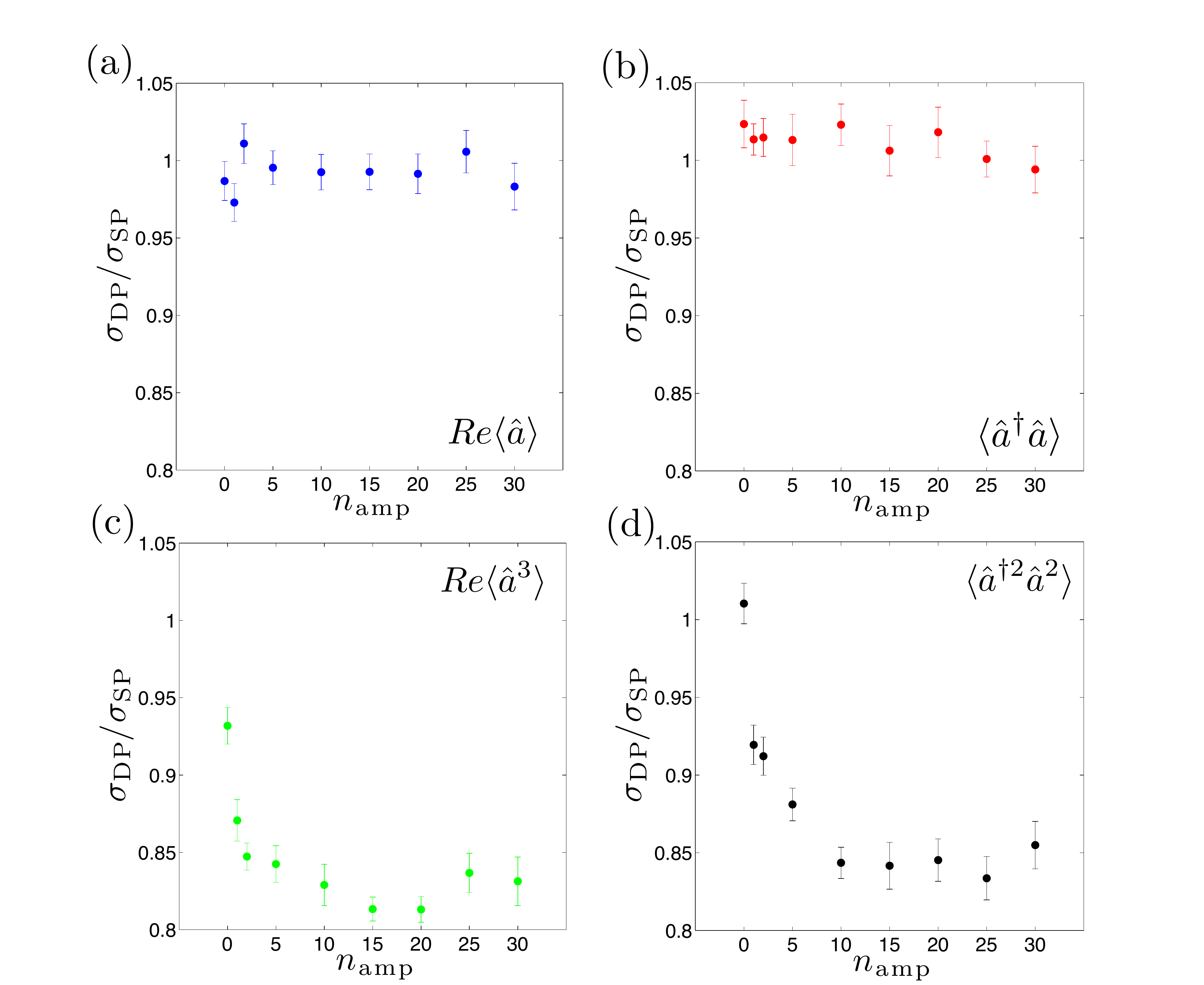}
\caption{Ratio of the standard deviations for the reconstruction of the DPM ($\sigma_{\text{DP}}$) and the SPM ($\sigma_{\text{SP}}$) in dependence on the photon number $n_{\text{amp}}$ of the amplifier noise fields.}
\label{Figure4}
\end{figure}

\begin{figure}[h!]
\centering
\includegraphics[width=1.03\textwidth]{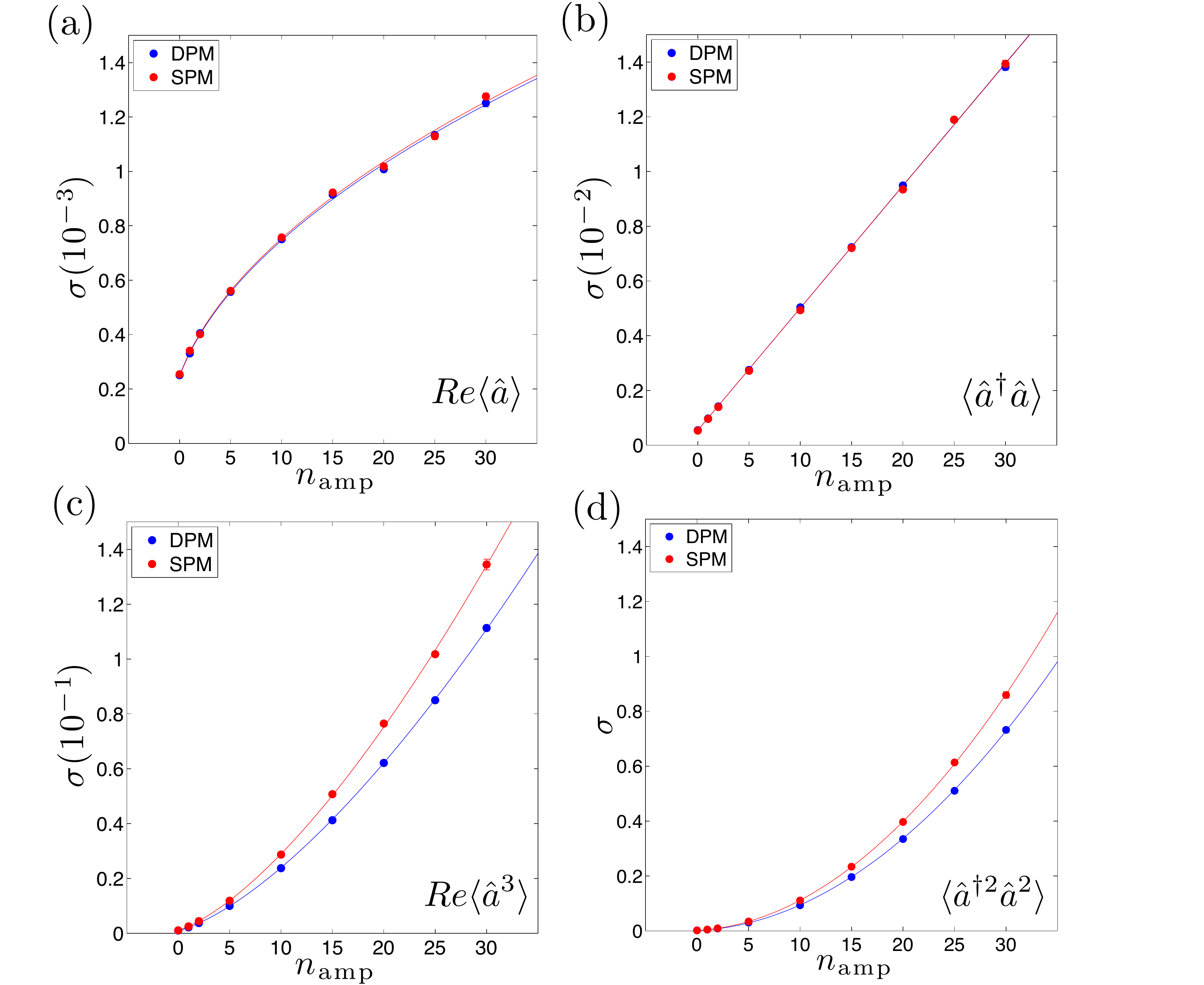}
\caption{Standard deviations ($\sigma$) for the reconstruction using the DPM and the SPM for different amplifier photon noises ($n_{\text{amp}}$). The results are fitted with functions $a\,(n_{\text{amp}}+b)^{k/2}$, where $k$ is the order of the moment and $a$, $b$ are the fitting coefficients.}
\label{Figure5}
\end{figure}

As second numerical test, we fix the number of measurements to $N=10^7$, and we simulate the two methods with different $n_{\text{amp}}$. Throughout this work, we assume that $n_{\text{amp}}$ is the same for all amplifiers. In Figures~\ref{Figure4}(c),(d) and ~\ref{Figure5}(c),(d), we observe that the DPM performs better than the SPM for the third and fourth moments for high $n_{\text{amp}}$, while for $n_{\text{amp}}=0$ both methods have approximately the same performance. From the good agreement between the polynomial fit curves and the data in Fig.~\ref{Figure5}, we conclude that the standard deviation depends on the amplifier noise as $\sigma\propto n_{\text{amp}}^{k/2}$ for large $n_{\text{amp}}$, where $k$ is the order of the considered moment. The exponent $k/2$ has also been observed in simulations of an early version of the DPM for sinusoidal signals~\cite{Summer1}. 

\section{Characterization of the output state}

In addition to quantum tomography of the input state, the DPM also allows one to obtain the output state of the beam splitter by simply using the beam splitter relations. Moreover, if the input signal of the beam splitter shows nonclassical behaviour, then the output can be entangled~\cite{Kim1}. Such entanglement can, in principle, be calculated from the reconstructed output state. However, it is desirable to measure the entanglement in the output state independently from a verification of the nature of the input state. Hence, we study a reliable scheme to directly detect entanglement between the two output signals, by making minimal assumptions on the beam splitter~\cite{Menzel2}. Indeed, we retrieve the moments of the output state by extending the reference-state method introduced in section~\ref{referencestate} to a two-path situation. Then, one can use entanglement witnesses or criteria based on moments~\cite{Vogel1, Miranowicz1, Miranowicz2} and, under certain reasonable assumptions, quantify the entanglement by using an entanglement measure~\cite{Horodecki1}. 

\subsection{Dual-path reference-state method}

We want to check and estimate the entanglement between the output modes of the beamsplitter $\hat s_1$ and $\hat s_2$, by minimal assumptions on the beam splitter. To achieve this goal, we can perform a reference-state method adapted for two channels. First, we send a known state to characterize the correlations between the two amplifier noise fields $\langle\hat V_1^{k_1}\hat V_1^{\dag j_1}\hat V_2^{k_2}\hat V_2^{\dag j_2}\rangle$ by making use of the known moments of the reference state $\langle\hat s_1^{\dag l_1}\hat s_1^{m_1}\hat s_2^{\dag l_2}\hat s_2^{m_2}\rangle$, and algebraically inverting the following equation
\begin{align}
   \label{Sonepath}
   \langle S_1^{\dag l_1}S_1^{m_1}
   S_2^{\dag l_2}S_2^{m_2}\rangle &  =
    \sum_{k_1=0}^{l_1}\sum_{k_2=0}^{l_2}
    \sum_{j_1=0}^{m_1}\sum_{j_2=0}^{m_2}
    {l_1\choose k_1}{l_2\choose k_2}
    {m_1\choose j_1}{m_2\choose j_2}
    \nonumber\\
    \quad&\times\langle
    \hat s_1^{\dag l_1-k_1}\hat s_1^{m_1-j_1}
    \hat s_2^{\dag l_2-k_2}\hat s_2^{m_2-j_2}
    \rangle
      \langle
    \hat V_1^{k_1} \hat V_1^{\dag j_1}
    \hat V_2^{k_2} \hat V_2^{\dag j_2}
    \rangle\,.
\end{align}
Here, we have used the complex envelopes defined in \eqref{envelope}:
\begin{align}\label{ref1}
\hat S_1&=\hat s_1+\hat V_1^\dag,\\
\hat S_2&=\hat s_2+\hat V_2^\dag.
\end{align}
We note that by measuring correlations, we can also evaluate whether the two amplifier noise contributions are correlated or not, while in the DPM it was assumed that they are uncorrelated. In the case of vacuum as inputs of the beam splitter, which is the most simple choice in the case of microwave experiments at millikelvin temperatures, we have simply
\begin{align}\label{ref2}
\langle\hat s_1^{\dag l_1-k_1}\hat s_1^{m_1-j_1}\hat s_2^{\dag l_2-k_2}\hat s_2^{m_2-j_2}\rangle_{\text{known}}=
\delta_{l_1,k_1}\delta_{m_2,j_2}\delta_{m_1,j_1}\delta_{,l_2,k_2}.
\end{align}
Then, we send the desired input signal and measure the correlations between $\hat S_1$ and $\hat S_2$. Inverting~\eqref{Sonepath} with respect to $\langle\hat s_1^{\dag l_1-k_1}\hat s_1^{m_1-j_1}\hat s_2^{\dag l_2-k_2}\hat s_2^{m_2-j_2}\rangle$, we find auto- and cross-correlations between the two modes $\hat s_1$ and $\hat s_2$. We note that we have not used any details of the beam splitter, such as input-ouput relations, in the derivation of Eqs.~\eqref{ref1}-\eqref{ref2} and thus treat it as a black box. We only assume that if the input states are vacuum states, also the output is a vacuum state. This assumption is well justified since a cold beam splitter is a passive device containing no energy sources~\cite{Hoffmann1, Collins1}.

\subsection{Entanglement detection}

Once we have measured the correlations of the two output modes, we can use criteria based on moments to check if there is entanglement. These kind of criteria have been well investigated in~\cite{Vogel1, Miranowicz1, Miranowicz2}. As example, we consider an entanglement witness based on the entanglement of Gaussian states.
Another issue is how to quantify entanglement starting from the measured moments. There are different entanglement measures based on different features, but all of them obey some basic properties~\cite{Horodecki1}. The estimation of an entanglement measure for an arbitrary state is still an open problem. Analytical solutions have been proposed for some classes of states, e.g., Gaussian states~\cite{Adesso1} (see section~\ref{examples}). Furthermore, techniques have been proposed~\cite{Eisert1, Guhne1} to determine a lower bound for the degree of entanglement even in the case when only a finite number of moments is available (incomplete tomography). Usually, they require the solution of a convex optimization problem, that can be done efficiently if the dimension of the Hilbert space is not too large. For instance, let us define the negativity~\cite{Vidal1}
\begin{equation}
\mathcal{N}(\rho)\equiv\frac{\|\rho^{T_1}\|_1-1}{2},
\end{equation}
where $\rho$ is the density matrix of the two-mode state after the beam splitter and $\rho^{T_1}$ denotes the partial transpose with respect to the system $1$. A lower bound on $\mathcal{N}(\rho)$ is given by solving the minimization problem
\begin{align}\label{Neg}
\text{minimize} \quad&\mathcal{N}(\sigma)\\
\text{subject to}\quad & \langle\hat s_1^{\dag k_1}\hat s_1^{k_2}\hat s_2^{\dag k_3}\hat s_2^{k_4}\rangle_{\sigma}=\langle\hat s_1^{\dag k_1}\hat s_1^{k_2}\hat s_2^{\dag k_3}\hat s_2^{k_4}\rangle_{\rho}    \nonumber\\
\quad&\sigma\succeq0\nonumber\\
\quad& \text{Tr}\,\sigma=1, \nonumber
\end{align}
that can be reshaped as a semidefinite program~\cite{Eisert1}.

Finally, we emphasize that the proposed detection method respects some basic criteria for a reliable experimental entanglement verification~\cite{vanEnk1}. In particular, it  does not make any assumption on the input state and it is independent of the state generation process. 

\section{Examples}

The DPM has been applied in an experiment to reconstruct the Wigner function of squeezed states~\cite{Menzel2}. In the same work (see also Ref.~\cite{Flurin1} for an alternative method), the authors use the dual-path setup to generate and quantify spatially separated entanglement, by sending a squeezed state and a vacuum to a microwave beam splitter~\cite{Mariantoni2, Hoffmann1}. In this section, we present two experimental examples, corresponding to similar data. In the first one, we perform quantum tomography of the input state, assuming the beam splitter input-output relations. In the second one, we detect entanglement of the beam splitter output making the black box assumption. For this purpose, we use an entanglement witness based on the analytical solution of the negativity for Gaussian states~\cite{Adesso1}. 

\subsection{Experimental reconstruction of a single squeezed vacuum state}
\begin{figure}[t]
\centering
\includegraphics[width=1.\textwidth]{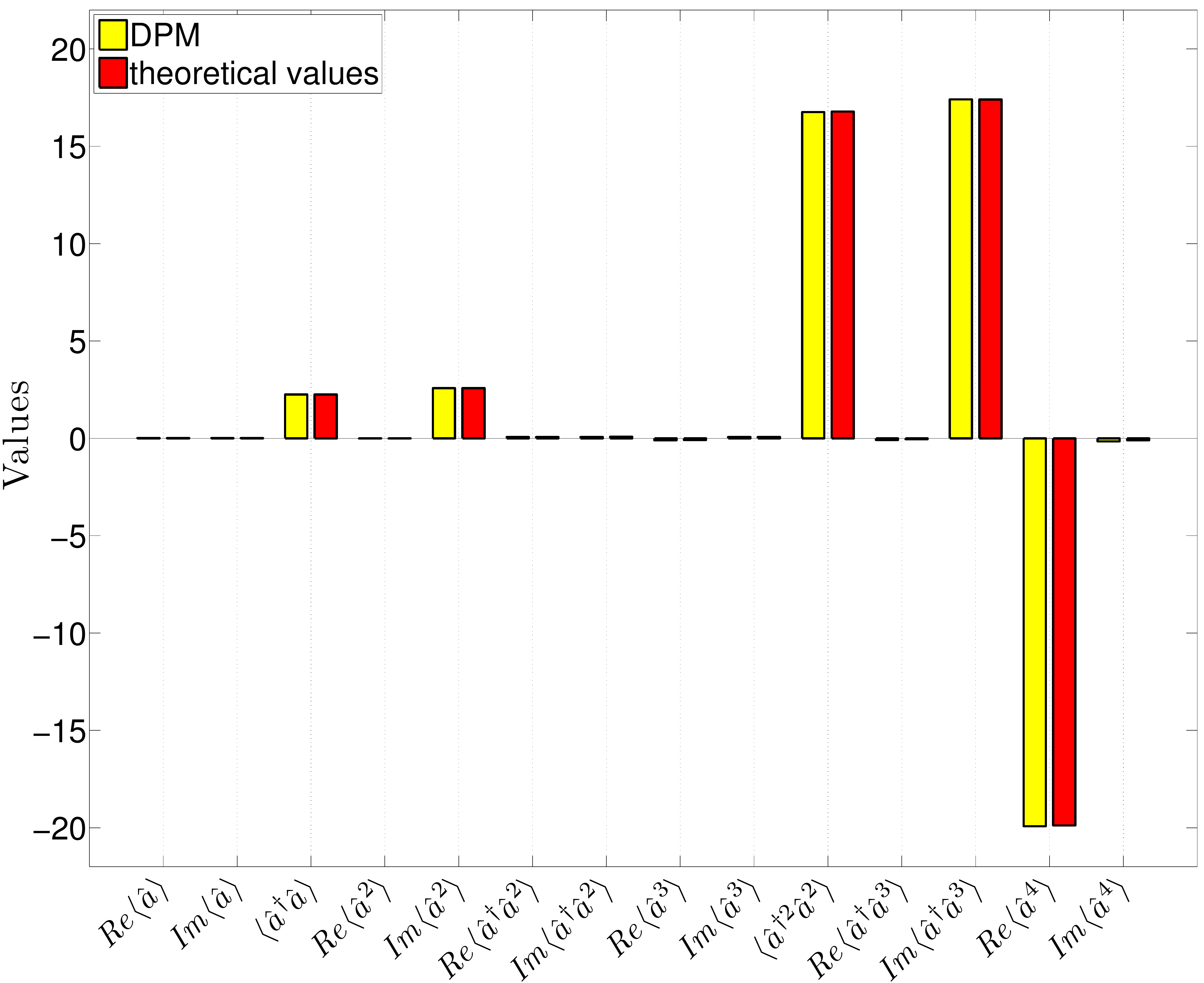}
\caption{Values of the reconstructed moments of the input signal $\hat a$ versus the theoretical values. The error bar is in the linewidth.}
\label{Figure8}
\end{figure}
In the experiment, we send a squeezed vacuum state and a vacuum state to a beam splitter. The squeezed state is generated by sending the vacuum signal produced by a $50\,\Omega$-resistor to a JPA operated in the degenerate mode.  The operation point of the JPA is characterized by a non-degenerate signal gain of $5.1$ dB. The two outputs of the beam splitter are amplified via two HEMT amplifiers working at 4 K. The measurements are performed using two IQ mixers working at room temperature. The data processing is realized using an FPGA logic~\cite{Menzel2}. The number of the measurements registered is $\sim 5\times10^9$. In Figure~\ref{Figure8}, we represent the reconstructed moments for the input signal $\hat a$.
We compare these moments with the ones of the Gaussian state defined by the first two reconstructed moments. We see that the moments correspond very well to the ones expected for Gaussian states. We note that the reconstructed squeezed state is not pure, because we have $\Delta \hat x_{\theta}^2=0.179\pm0.001$ and $\Delta \hat x_{\theta+\frac{\pi}{2}}^2=5.326\pm0.006$, where $\hat x_{\theta}$ is the quadrature with minimum  variance, and these values lead to $\Delta \hat x_{\theta} \Delta \hat x_{\theta+\frac{\pi}{2}}= 0.978\pm0.006>0.5$. For this reason, four moments are not enough for full-tomography, so we simply check the Gaussianity of the state by comparing the third and fourth moment with their theoretical values (see Fig.~\ref{Figure8}). The squeezing obtained is  $4.45 \pm0.03$ dB with respect to the variance of the vacuum state~\cite{Menzel2}, and this is a witness of non-classicality of the state~\cite{Vogel2} (we note that the squeezing depends only on the first two moments, which are very well estimated). The moments of the amplifier noise operators agree with a thermal state with an average photon number $12.239\pm0.002$ and $12.805 \pm 0.002$, similar to the value used for the simulations in section~\ref{comparison}. 

\subsection{Experimental detection of a path-entangled continuous-variable microwave state}\label{examples}

In the next example, we use the extended reference-state method to detect the entanglement of the output of the beam splitter, in the same experimental conditions as in the previous example. As we want to prove that the beam splitter is an entangler, we do not assume its input-output relations. In this sense, we need first to send vacuum as input to reconstruct the HEMT amplifier noise, and, then, we measure the output moments sending a squeezed state as input. We can reconstruct the moments of the beam splitter output by using Eq.~\eqref{Sonepath}, and by using witnesses based on moments, we can assert whether there is entanglement. The witness that we have used is based on the analytical solution of the negativity for a two-mode Gaussian state, given by~\cite{Adesso1}
\begin{equation}\label{Adesso}
\mathcal{N}(\rho_G)=\max\left\{0,\frac{1-\nu}{2\nu}\right\}\equiv\max\left\{0,\widetilde{ \mathcal{N}_K}\right\},
\end{equation}
where $\rho_G$ indicates a Gaussian state, and $\nu$ is the smallest symplectic value of the two-mode Gaussian state~\cite{Adesso1} (see supp. material of~\cite{Menzel2} for an explicit expression). Here, the negativity kernel $\widetilde {\mathcal{N}_K}$~\cite{Menzel2} is an entanglement witness, given that if a non-Gaussian state has the same first two moments as an entangled Gaussian state, then it is entangled~\cite{Hyllus1}. If we assume the Gaussianity of the state, as it is in our case, then $\widetilde {\mathcal{N}_K}$ is directly related to an entanglement measure via Eq.~\eqref{Adesso}. By using the same sample data of the previous example, we obtain a negativity of $0.489\pm0.004$.

\section{Summary and conclusions}

We have studied the quantum aspects of the dual-path state reconstruction method introduced in Ref.~\cite{Menzel1}, giving explicit formulas for the reconstruction of the field quadrature moments. Our numerical simulations show that the DPM performs better than the SPM introduced in Ref.~\cite{Eichler1, Eichler2} at the cost of some experimental resources: an additional beam splitter and an additional amplifying channel. We have investigated the entanglement generated at the output of the beam splitter by proposing a detection scheme that does not assume the knowledge of beam splitter relations. Furthermore, we have experimentally tested the method with a squeezed state and vacuum as input fields of the beam splitter, showing that the method works in a realistic situation. In conclusion, we have proposed a toolbox for quantum information with propagating quantum microwaves, consisting of different quantum tomographic and entanglement detection schemes. With respect to a possible application of our dual-path tomography in quantum information architectures, scaling to multiple modes is straightforward in two important cases. First, for frequency multiplexing, a single beam splitter is sufficient and only the processing of its two output paths needs to be changed (e.g. multichannel digital down conversion and parallel processing). Second, for the analysis of signals consisting of multiple spatially separate modes, each mode needs its own beam splitter. In this way, one obtains the noise moments of each output path in analogy to the single-mode case and the multi-mode state can be reconstructed. All in all, our results may pave the way for future microwave quantum teleportation and microwave quantum communication developments.

\section*{Acknowledgments}
We thank D. Ballester for useful discussions and feedback on this work. The authors acknowledge support from Spanish MINECO FIS2012-36673-C03-02; UPV/EHU UFI 11/55; Basque Government IT472-10; SOLID, CCQED, PROMISCE, and SCALEQIT European projects; the Deutsche Forschungsgemeinschaft via SFB 631; the German Excellence Initiative via the `Nanosystem Initiative Munich' (NIM).

\section*{Appendix: Dual-path formulas}
Here, we write the general formulas to retrieve the moments of an input field $\langle\hat a^{\dag l}\hat a^m\rangle$ ($l+m\geq1$), the moments of the noise in the channels $\langle\hat V_{1,2}^r\hat V_{1,2}^{\dag s}\rangle$ ($r+s\geq2$), and the moments of the ancilla $\langle\hat v^{\dag l}\hat v^m\rangle$ ($l+m\geq3$), from the observable moments of the outcoming signal $\langle\hat S_1^{\dag l_1}\hat S_2^{\dag l_2}\hat S_2^{m_2}\hat S_1^{m_1}\rangle$, with $l_1,l_2,m_1,m_2\geq0$. The formulas are recursive, in the sense that the higher-order moments depend on the lower-order ones.

From Eq.~\eqref{S2} we derive the general expression for the observable moments of the outcoming signals

\begin{align}\label{momS}
 &\langle\hat S_1^{\dag l_1}\hat S_2^{\dag l_2}\hat S_2^{m_2}\hat S_1^{m_1}\rangle=\sum_{k_1=0}^{l_1}\sum_{k_2=0}^{l_2}\sum_{j_1=0}^{m_1}\sum_{j_2=0}^{m_2}\sum_{k'_1=0}^{l_1-k_1}\sum_{k'_2=0}^{l_2-k_2}\sum_{j'_1=0}^{m_1-j_1}\sum_{j'_2=0}^{m_2-j_2}\nonumber\\
&\quad\quad\binom{l_1}{k_1}\binom{l_2}{k_2}\binom{m_1}{j_1}\binom{m_2}{j_2}\binom{l_1-k_1}{k'_1}\binom{l_2-k_2}{k'_2}\binom{m_1-j_1}{j'_1}\binom{m_2-j_2}{j'_2}\nonumber\\
&\quad\quad 2^{-(k_1+k'_1+j_1+j'_1+k_2+k'_2+j_2+j'_2)/2}(-1)^{j_2+k_2}\langle\hat a^{\dag k_1+k_2}\,\hat a^{j_1+j_2}\rangle\langle\hat v^{\dag k'_1+k'_2}\,\hat v^{j'_2+j'_1}\rangle\nonumber\\
&\quad\quad \langle\hat V_1^{l_1-k_1-k'_1}\,\hat V_1^{\dag m_1-j_1-j'_1}\hat V_2^{l_2-k_2-k'_2}\hat V_2^{\dag m_2-j_2-j'_2}\rangle,
\end{align} 
where we have used the binomial theorem and the fact that different modes commute with each others and that the noise fields are independent from the signal. In the following we will assume that the two noise fields $\hat V_1$ and $\hat V_2$ are independent from each others. At this point, we have a series of equations, each of which corresponds to a different choice of $l_1$, $l_2$, $m_1$, $m_2$. Solving this system of equations, we find the required formulas. A possible solution for the moments of $\hat a$ is 
\begin{align}
\langle\hat a\rangle&=2^{-1/2}(\langle\hat S_1\rangle-\langle\hat S_2\rangle),\\
\langle\hat a^2\rangle&=-2\langle\hat S_1\hat S_2\rangle+\langle\hat v^2\rangle,\\
\langle\hat a^\dag  \hat a\rangle&=-(\langle\hat S_1^\dag\hat S_2\rangle+\langle\hat S_2^\dag\hat S_1\rangle)+\langle\hat v^{\dag}_{} \hat v_{}^{}\rangle,\\
\langle\hat a^{\dag l}_{}\hat a_{}^{m}\rangle&=\frac{1}{lm+m+l-1}\sum_{l_1=0}^l\sum_{m_1=0}^m\langle \hat a_{}^{\dag l}\hat a_{}^{m}\rangle_{l_1m_1}(1-\delta_{l_1,l}\delta_{m_1,m}-\delta_{l_1,0}\delta_{m_2,0})\nonumber\\
\quad\quad&\qquad\qquad\qquad\qquad\qquad\qquad\qquad\qquad\qquad\qquad l+m>2.
\end{align}
Here, $\langle\hat a^{\dag l}\hat a^m\rangle_{l_1m_1}$ is the algebraic inversion of \eqref{momS} with $m_2=m-m_1$ and $l_2=l-l_1$
\begin{align}\label{a}
&\langle(\hat a^\dag)^l\hat a^{m}\rangle_{l_1,m_1}=(-1)^{l-l_1+m-m_1}2^{(l+m)/2}\langle\hat S_1^{\dag l_1}\hat S_2^{\dag l-l_1}\hat S_2^{m-m_1}\hat S_1^{m_1}\rangle\nonumber\\
&\quad-\sum_{k_1=0}^{l_1}\sum_{k_2=0}^{l-l_1}\sum_{j_1=0}^{m_1}\sum_{j_2=0}^{m-m_1-1}\sum_{k'_1=0}^{l_1-k_1}\sum_{k'_2=0}^{l-l_1-k_2}\sum_{j'_1=0}^{m_1-j_1}\sum_{j'_2=0}^{m-m_1-j_2}\binom{l_1}{k_1}\binom{l-l_1}{k_2}\binom{m_1}{j_1}\nonumber\\
&\quad\quad\binom{m-m_1}{j_2}\binom{l_1-k_1}{k'_1}\binom{l-l_1-k_2}{k'_2}\binom{m_1-j_1}{j'_1}\binom{m-m_1-j_2}{j'_2}\,\nonumber\\
&\quad\quad (-1)^{l-l_1+m-m_1+j_2+k_2}2^{(l+m-k_1-k'_1-j_1-j'_1-k_2-k'_2-j_2-j'_2)/2}\langle\hat a^{\dag k_1+k_2}\,\hat a^{j_1+j_2}\rangle\nonumber\\
&\quad\quad \langle\hat v^{\dag k'_1+k'_2}\,\hat v^{j'_2+j'_1}\rangle\langle\hat V_1^{l_1-k_1-k'_1}\,\hat V_1^{\dag m_1-j_1-j'_1}\rangle\langle\hat V_2^{l-l_1-k_2-k'_2}\hat V_2^{\dag m-m_1-j_2-j'_2}\rangle \nonumber\\
&\quad-\sum_{k_1=0}^{l_1}\sum_{k_2=0}^{l-l_1}\sum_{j_1=0}^{m_1-1}\sum_{k'_1=0}^{l_1-k_1}\sum_{k'_2=0}^{l-l_1-k_2}\sum_{j'_1=0}^{m_1-j_1}\binom{l_1}{k_1}\binom{l-l_1}{k_2}\binom{m_1}{j_1}\binom{l_1-k_1}{k'_1}\nonumber\\
&\quad\quad \binom{l-l_1-k_2}{k'_2}\binom{m_1-j_1}{j'_1}(-1)^{l-l_1+k_2} 2^{(l+m_1-k_1-k'_1-j_1-j'_1-k_2-k'_2)/2}\nonumber\\
&\quad\quad \langle\hat a^{\dag k_1+k_2}\,\hat a^{j_1+m-m_1}\rangle\langle\hat v^{\dag k'_1+k'_2}\,\hat v^{j'_1}\rangle\langle\hat V_1^{l_1-k_1-k'_1}\,\hat V_1^{\dag m_1-j_1-j'_1}\rangle\langle\hat V_2^{l-l_1-k_2-k'_2}\rangle \nonumber\\
&\quad-\sum_{k_1=0}^{l_1}\sum_{k_2=0}^{l-l_1-1}\sum_{k'_1=0}^{l_1-k_1}\sum_{k'_2=0}^{l-l_1-k_2}\binom{l_1}{k_1}\binom{l-l_1}{k_2}\binom{l_1-k_1}{k'_1}\binom{l-l_1-k_2}{k'_2}(-1)^{l-l_1+k_2}\,\nonumber\\
&\quad\quad 2^{(l-k_1-k'_1-k_2-k'_2)/2}\langle\hat a^{\dag k_1+k_2}\,\hat a^{m}\rangle\langle\hat v^{\dag k'_1+k'_2}\rangle\langle\hat V_1^{l_1-k_1-k'_1}\rangle\langle\hat  V_2^{l-l_1-k_2-k'_2}\rangle\nonumber\\
&\quad-\sum_{k_1=0}^{l_1-1}\sum_{{k'}_1=0}^{l_1-k_1}\binom{l_1}{k_1}\binom{l_1-k_1}{k'_1}\,2^{(l_1-k_1-k'_1)/2}\langle\hat a^{\dag k_1+l-l_1}\,\hat a^{m}\rangle\langle\hat v^{\dag k'_1}\rangle\langle\hat V_1^{l_1-k_1-k'_1}\rangle.
\end{align}
The moments of the channel noise fields $\hat V_1$ and $\hat V_2$ are 
\begin{align}
&\langle\hat V_1\rangle=\langle\hat V_2\rangle=0,\\
&\langle\hat V_1^r\hat{V}_1^{\dag s}\rangle=\langle\hat S_1^{\dag r}\hat S_1^s\rangle\nonumber\\
\quad&-\sum_{k_1=0}^{r}\sum_{j_1=0}^{s-1}\sum_{k'_1=0}^{r-k_1}\sum_{j'_1=0}^{s-j_1}\binom{r}{k_1}\binom{s}{j_1}\binom{r-k_1}{k'_1}\binom{s-j_1}{j'_1}\,2^{(k_1+j_1-r-s)/2}\langle\hat a^{\dag k'_1}\,\hat a^{j'_1}\rangle\nonumber\\
&\quad\quad \langle\hat v^{\dag r-k_1-k'_1}\,\hat v^{s-j_1-j'_1}\rangle\langle\hat V_1^{k_1}\hat V_1^{\dag j_1}\rangle \nonumber\\
&\quad-\sum_{k_1=0}^{r-1}\sum_{k'_1=0}^{r-k_1}\binom{r}{k_1}\binom{r-k_1}{k'_1}\,2^{(k_1-r)/2}\langle\hat a^{\dag k'_1}\rangle\langle\hat v^{\dag r-k_1-k'_1}\rangle\langle\hat V_1^{k_1}\hat V_1^{\dag s}\rangle,\label{V1}\\
&\langle\hat V_2^r\hat V_2^{\dag s}\rangle=\langle\hat S_2^{\dag r}\hat S_2^s\rangle\nonumber\\
&\quad-\sum_{k_1=0}^{r}\sum_{j_1=0}^{s-1}\sum_{k'_1=0}^{r-k_1}\sum_{j'_1=0}^{s-j_1}\binom{r}{k_1}\binom{s}{j_1}\binom{r-k_1}{k'_1}\binom{s-j_1}{j'_1}(-1)^{k'_1+j'_1}\,2^{(k_1+j_1-r-s)/2}\nonumber\\
&\quad\quad \langle\hat a^{\dag k'_1}\,\hat a^{j'_1}\rangle\langle\hat v^{\dag r-k_1-k'_1}\,\hat v^{s-j_1-j'_1}\rangle\langle\hat V_2^{k_1}\hat V_2^{\dag j_1}\rangle \nonumber\\
&\quad-\sum_{k_1=0}^{r-1}\sum_{k'_1=0}^{r-k_1}\binom{r}{k_1}\binom{r-k_1}{k'_1}(-1)^{k'_1}\,2^{(k_1-r)/2}\langle\hat a^{\dag k'_1}\rangle\langle\hat v^{\dag r-k_1-k'_1}\rangle\langle\hat V_2^{k_1}\hat{V}_2^{\dag s}\rangle.\label{V2}
\end{align}
And, the moments of the ancilla $\hat v$ are
\begin{align}
\langle\hat v^{\dag l}_{}\hat v_{}^{m}\rangle&=\frac{1}{lm+m+l-1}\sum_{l_1=0}^l\sum_{m_1=0}^m\langle \hat v_{}^{\dag l}\hat v_{}^{m}\rangle_{l_1m_1}(1-\delta_{l_1,l}\delta_{m_1,m}-\delta_{l_1,0}\delta_{m_2,0})\nonumber\\
\quad\quad&\qquad\qquad\qquad\qquad\qquad\qquad\qquad\qquad\qquad\qquad l+m>2.
\end{align}
where $\langle \hat v^{\dagger l}\hat v^m\rangle_{l_1,m_1}$ is given by Eq.~\eqref{a} after replacing: $\hat v\rightarrow \hat a$ ($\hat v^\dag\rightarrow \hat a^\dag$) as well as $(-1)^{l-l_1+m-m_1}\rightarrow 1$, $(-1)^{l-l_1+m-m_1+j_2+k_2}\rightarrow (-1)^{j'_2+k'_2}$, and $(-1)^{l-l_1+k_2}\rightarrow (-1)^{k'_2}$.
\\
\\
\\


\section*{References}

\begin{thebibliography}{99}

\bibitem{Blais1} Blais A, Huang R-S, Wallraff  A, Girvin S M, and Schoelkopf R J 2004 {\it Phys. Rev. A} {\bf 69} 062320

\bibitem{Wallraff1} Wallraff A, Schuster D I, Blais A, Frunzio L, Huang R-S, Majer J, Kumar S, Girvin S M, and Schoelkopf R J 2004 {\it Nature} {\bf 431} 162

\bibitem{Braunstein1} Braunstein S L, van Loock P 2005 {\it Rev. Mod. Phys.} {\bf 77} 513

\bibitem{Wineland1} Leibfried D, Meekhof D M, King B E, Monroe C, Itano W M, and Wineland D J 1996 {\it Phys. Rev. Lett.} {\bf 77} 4281

\bibitem{Schiller1} Lvovsky A I, Hansen H, Aichele T, Benson O, Mlynek J, and Schiller S 2001 {\it Phys. Rev. Lett.} {\bf 87} 050402

\bibitem{Solano1} Lougovski P, Solano E, Zhang Z M, Walther H, Mack H, and Schleich W P 2003 {\it Phys. Rev. Lett.} {\bf 91} 010401

\bibitem{Haroche1} DelŽglise S, Dotsenko I, Sayrin C, Bernu J, Brune M, Raimond J-M, and Haroche S 2008 {\it Nature} {\bf 455} 510

\bibitem{Martinis1} Hofheinz M, Wang H, Ansmann M, Bialczak R C, Lucero E, Neeley M, O'Connell A D, Sank D, Wenner J, Martinis J M, and Cleland A N 2009 {\it Nature} {\bf 459} 546

\bibitem{Romero1} Romero G,  Garcia-Ripoll J J, and Solano E 2009 {\it Phys. Rev. Lett.} {\bf 102} 173602

\bibitem{Peropadre1} Peropadre B, Romero G, Johansson G, Wilson C M, Solano E, and Garcia-Ripoll J J 2011 {\it Phys. Rev. A} {\bf 84} 063834

\bibitem{Chen1} Chen Y-F, Hover D, Sendelbach S, Maurer L, Merkel S T, Pritchett E J, Wilhelm F K, and McDermott R 2011 {\it Phys. Rev. Lett.} {\bf 107} 217401

\bibitem {Menzel1} Menzel E P, Deppe F, Mariantoni M, Araque Caballero M \'A, Baust A, Niemczyk T,  Hoffmann E,  Marx A, Solano E, and Gross R 2010 {\it Phys. Rev. Lett.} {\bf 105} 100401

\bibitem{Mariantoni1} Mariantoni  M, Storcz M J, Wilhelm F K, Oliver W D, Emmert A, Marx A, Gross R, Christ H, and Solano E 2005 {\it arxiv:cond-mat/0509737}

\bibitem{Menzel2} Menzel E P, Di Candia R, Deppe F, Eder P, Zhong L, Ihmig M, Haeberlein M, Baust A, Hoffmann E, Ballester D, Inomata K, Yamamoto T, Nakamura Y, Solano E, Marx A, and Gross R 2012 {\it Phys. Rev. Lett.} {\bf 109} 250502

\bibitem{Buzek1} Bu\v{z}ek V, Adam G, and Drobn\'y G 1996 {\it Phys. Rev. A} {\bf 54} 804; Bu\v{z}ek V, Adam G, and Drobn\'y G 1996 {Ann. Phys.} {\bf 245} 37

\bibitem{Yamamoto1} Yamamoto T, Inomata K, Watanabe M, Matsuba K, Miyazaki T, Oliver W D, Nakamura Y, Tsai J S 2008 {\it Appl. Phys. Lett.}  {\bf 93} 042510

\bibitem{Castellano1} Castellanos-Beltran M A, Irwin K D, Hilton G C, Vale L R, and Lehnert K W 2008 {\it Nature Phys.} {\bf 4} 929

\bibitem{Zhong1} Zhong L, Menzel E P, Di Candia R, Eder P, Ihmig M, Baust A, Haeberlein M, Hoffmann E, Inomata K, Yamamoto T, Nakamura Y, Solano E, Deppe F, Marx A, and Gross R 2013 {\it New J. Phys.} {\bf 12} 125013

\bibitem{Eichler1} Eichler C, Bozyigit D, Lang C, Steffen L, Fink J, and Wallraff A 2011 {\it Phys. Rev. Lett.} {\bf 106} 220503

\bibitem{Eichler2} Eichler C, Bozyigit D, Lang C, Baur M, Steffen L, Fink J M, Filipp S, and Wallraff A 2011 {\it Phys. Rev. Lett.} {\bf 107} 113601

\bibitem{Leonhardt1} Leonhardt U, and Paul H 1994  {\it Phys. Rev. Lett.} {\bf 72} 4086

\bibitem{Mallet1} Mallet F, Castellano-Beltran M A, Ku H S, Glancy S, Knill E, Irwin K D, Hilton G C, Vale L R, and Lehnert K W 2011 {\it Phys. Rev. Lett.} {\bf106} 220502

\bibitem{Vogel1} Shchukin E, and Vogel W 2005 {\it Phys. Rev. Lett.} {\bf 95} 230502.  Miranowicz A, and Piani M 2006 {\it Phys. Rev. Lett.} {\bf 97} 058901

\bibitem{Miranowicz1} Miranowicz A, Piani M, Horodecki P, and Horodecki R 2009 {\it Phys. Rev. A} {\bf 80} 052303

\bibitem{Miranowicz2} Miranowicz A, Bartkowiak M, Wang X, Liu Y-X, and Nori F 2010 {\it Phys. Rev. A} {\bf 82} 013824

\bibitem{Mariantoni2} Mariantoni M, Menzel E P, Deppe F, Araque Caballero M \'A, Baust A, Niemczyk T, Hoffmann E, Solano E, Marx A, and Gross R 2010 {\it Phys. Rev. Lett.} {\bf 105} 133601 

\bibitem{Hoffmann1} Hoffmann E, Deppe F, Niemczyk T, Wirth T, Menzel E P, Wild G, Huebl H, Mariantoni M, Wei{\ss}l T, Lukashenko A, Zhuravel A P, Ustinov A V, Marx A, and Gross R 2010 {\it Appl. Phys. Lett.}, {\bf 97} 222508

\bibitem{Caves1} Caves C M 1982 {\it Phys. Rev. D} {\bf 26} 1817

\bibitem{Collins1} Collin R E 2001 {\it Foundation for Microwave Engineering} Wiley-IEEE, New York

\bibitem{daSilva1} da Silva M P, Bozyigit D, Wallraff A, and Blais A 2010 {\it Phys. Rev. A } {\bf 82} 043804 

\bibitem{Friis} Friis H T, 1944 {\it Proc. IRE} {\bf 32} 419

\bibitem{Barnett} Barnett S, and Radmore P M. 2002 {\it Methods in theoretical quantum optics} Oxford University Press

\bibitem{Kim1} Kim M S, Son W, Bu\v{z}ek V, and Knight P L 2002 {\it Phys. Rev. A} {\bf 65} 032323

\bibitem{Summer1} Summer P, 2011 {\it Simulation of the dual-path detector for propagating quantum microwaves}  Bachelor thesis Technische Universit{\"{a}}t M{\"{u}}nchen, available online \url{www.wmi.badw-muenchen.de/publications/theses/Summer_Bachelorarbeit_2011.pdf}

\bibitem{Horodecki1} Horodecki R, Horodecki P, Horodecki M, Horodecki K 2009 {\it Rev. Mod. Phys.} {\bf 81} 865

\bibitem{Adesso1} Adesso G, and Illuminati F 2005 {\it Phys. Rev. A} {\bf 72} 032334

\bibitem{Eisert1} Eisert J, Brandao F G S L, and Audenaert K M R 2007 {\it New J. Phys.} {\bf 9} 46

\bibitem{Guhne1} G\"uhne O, Reimpell M, and Werner R F 2007 {\it Phys. Rev. Lett.} {\bf 98} 110502

\bibitem{Vidal1} Vidal G, and Werner R F 2002 {\it Phys. Rev. A} {\bf 65} 032314

\bibitem{vanEnk1} van Enk S J, L\"utkenhaus, N, and Kimble H J 2007 {\it Phys. Rev. A} {\bf 75} 052318

\bibitem{Flurin1} Flurin E, Roch N, Mallet F, Devoret M H, and Huard B 2012 {\it Phys. Rev. Lett.} {\bf 109} 183901

\bibitem{Vogel2} Vogel W 2000 {\it Phys. Rev. Lett.} {\bf 84} 1849 

\bibitem{Hyllus1} Hyllus P, and Eisert J 2006 {\it New. J. Phys.} {\bf 8} 51


\end{thebibliography}
\end{document}